\documentclass[journal=jctee,manuscript=article,layout=traditional]{achemso}
\pdfoutput=1
\usepackage{amsmath}
\usepackage{graphicx}
\usepackage{gensymb}
\usepackage{natbib}
\usepackage{cleveref}
\usepackage{subcaption}

\author{Devin J. Hernandez}
\affiliation{Department of Chemistry, University of California, Berkeley, California 94720, USA}

\author{Adam Rettig}
\affiliation{Department of Chemistry, University of California, Berkeley, California 94720, USA}

\author{Martin Head-Gordon}
\affiliation{Department of Chemistry, University of California, Berkeley, California 94720, USA}
\alsoaffiliation{Chemical Sciences Division, Lawrence Berkeley National Laboratory, Berkeley, California 94720, USA}
\email{mhg@cchem.berkeley.edu}

\title{A New View of Density Corrected DFT: \\ 
Can One Get a Better Answer for a Good Reason?}

\begin{document}

\begin{abstract}
    Despite its widespread use, density functional theory (DFT) has several notable
areas of failure; perhaps the most well-studied of these failures is self-interaction error (SIE). Density corrected DFT (DC-DFT) was proposed as a potential solution to systems where SIE causes traditional DFT to fail. The Hartree-Fock (HF) density is then used for cases where
the DFT energy is suitable but the self-consistent density is erroneous. In this study,
we investigate the utility of the higher quality orbital optimized MP2 densities in DC-DFT for barrier heights and halogen bonded complexes. For functionals such as PBE and r$^2$SCAN, find that these densities yield worse results than the HF density due to favorable cancellation between the density-driven and functional-driven errors, confirming a recent study. Error decomposition reveals functional driven error, not density driven error, to be the primary cause of inaccuracy in DFT calculations where SIE is prominent. We
therefore advise caution when using HF-DFT, because the only rigorous way to remove large functional-driven errors in lower rungs of Jacob's ladder is by climbing to higher rungs that include exact exchange. We recommend that better functionals be improved by using a better density in SIE-sensitive cases. Examples support the value of this variant of DC-DFT. We also emphasize that DC-DFT potential energy surfaces have first derivative discontinuities at Coulson-Fischer points, in contrast to the second derivative discontinuities in SCF solutions.

\end{abstract}

\section*{Introduction}
Density functional theory (DFT)\cite{kohn1996density,capelle2006bird,perdew2009some,burke2013dft} has become the workhorse of modern quantum chemistry as a result of its low computational cost and adequate accuracy for most chemically relevant applications\cite{mardirossian2017thirty,goerigk2017look,martin2020empirical}. 
Kohn-Sham DFT is a formally exact theory, but the exact functional is not known and a wide variety of density functional approximations are used instead.
These approximations are typically very good in the vast majority of normal chemical applications, but break down in a certain subset of cases.

\begin{figure}
    \centering
    \includegraphics[width=0.6\textwidth]{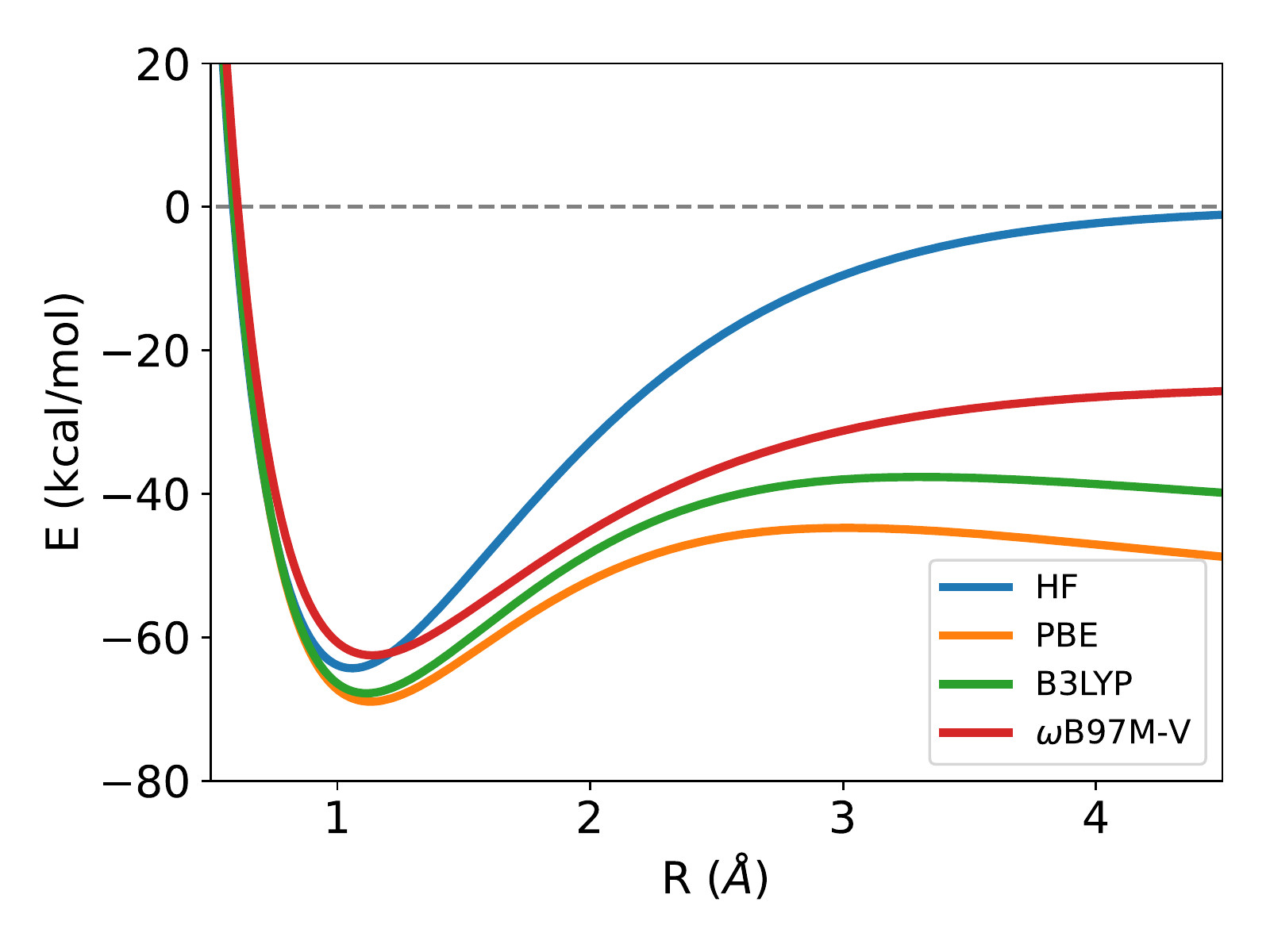}\\
    \caption{Dissociation of H$_2^+$  comparing 3 density functionals to Hartree-Fock (HF), which is exact in this case. PBE, B3LYP and $\omega$B97M-V all exhibit artificial energy lowering at long bond lengths. The hybrid B3LYP functional with 20\% exact exchange, removes only 20\% or so of the error, while the range separated hybrid, $\omega$B97M-V, with 100\% long-range exact exchange is better but still suffers from considerable error. For all three functionals, this density-driven SIE error is associated with 2 well-separated H$^{\frac{1}{2}+}$ atoms which are not properly degenerate with H + H$^+$, as in exact quantum mechanics.    }
    \label{fig:h2diss}
\end{figure}

Perhaps the most well studied of these breakdowns is self-interaction error (SIE) (also often called the delocalization error)\cite{polo2002electron,bao2018self, lundberg2005quantifying}, where fractional orbital occupation causes catastrophic failures in many systems including charge transfer states\cite{ruiz1996charge, zhang1998challenge}, conjugated systems, semiconductor band gaps, and organic cocrystals\cite{johnsondelocalization}.
SIE can be attributed to the Coulomb interaction between an electron with itself, causing the self-consistent electron density to artificially delocalize.\cite{cohen2008insights}
This is most apparent in systems such as H$_2^+$ where bond stretching predicts dissociation to the un-physically low energy limit of H$^{1/2+} \cdots $H$^{1/2+}$ demonstrated in \cref{fig:h2diss}. Other well-known examples are fractional charges in systems such as dissociating alkali-halide diatomics.\cite{dutoi2006self,ruzsinszky2006spurious}
Numerous solutions have been proposed to ameliorate this error, such as the Perdew-Zunger self-interaction correction which makes a density functional exact for one electron systems and adds no correction were the exact functional to be used\cite{PZSIC1981,perdew2015paradox}. Efforts to analyze and characterize SIE continue,\cite{hait2018delocalization} and SIE has recently been described as ``the greatest outstanding challenge in density functional theory.''\cite{johnsondelocalization}

Today, the most popular method to reduce SIE is to replace a fraction of the exchange functional with Hartree-Fock (HF) exchange also known as ``exact" exchange, as HF exchange cancels out the self-interaction present in the coulomb term of the total energy.\cite{becke1993new,becke1993thermo} These hybrid functionals have proved extremely successful\cite{raghavachari2000perspective} in improving the overall accuracy of DFT and partially correcting the SIE; however, in order to fully cancel out the SIE, we must replace 100$\%$ of the inexact exchange contributions. Development of functionals that employ 100\% exact exchange has been far more limited so far, although Becke has reported significant progress\cite{Hybridperspective2014,becke2022density}. A more common alternative are the range-separated functionals\cite{iikura2001long,gerber2005hybrid, baer2010tuned}, where the amount of exact exchange increases to 100$\%$ in the long range limit, with a tunable amount of exact exchange used in the short-range. These range-separated hybrid functionals proved even more accurate than global hybrid functionals,\cite{chai2008long,Mardirossian:2014,Mardirossian:2016} as well as reproducing the correct asymptotic behavior. Another alternative that is attracting interest are ``local hybrid'' functionals\cite{maier2019local,janesko2021replacing,furst2023full} where the fraction of exact exchange is position-dependent through features of the density.

Hybrid DFT achieved widespread success by replacing portions of the approximate exchange functional with HF exchange; one may take this even a step further by evaluating the DFT energy using the HF density, termed HF-DFT \cite{Pople1992, janesko2008hartree, verma2012increasing}. HF-DFT was introduced by the Pople group, before their early DFT code could self-consistently optimize the energy. This approach is currently motivated by the assumption that the DFT energy is accurate given a reasonable density, but self-consistent optimization of the energy can lead to qualitatively incorrect densities in SIE-sensitive cases. One therefore utilizes HF to produce a qualitatively correct density and evaluates the DFT energy non-self-consistently.

HF-DFT has been recognized as performing well for  difficult barrier height reaction problems,\cite{janesko2008hartree,verma2012increasing} but performance on databases such as GMTKN55\cite{goerigk2017look} was shown to be worse on average\cite{santradcdftgmtkn55}. This is perhaps unsurprising as many problems do not suffer significant SIE and have no need to be ``corrected"; additionally, HF does not always yield qualitatively correct densities due to its tendency to artificially break spin symmetry due to lack of correlation.  HF densities are also know to be slightly ``too ionic'' as neglect of electron correlation leads to dipole moments that are typically too large,\cite{bak2000accuracy,hickey2014benchmarking} and good density functionals perform considerably better.\cite{hait2018accurate} Likewise the HF density, as measured by its second moment, is a little too diffuse.\cite{hait2021too}

To reflect considerations such as those given above, Kim and coworkers\cite{kim2013understanding} introduced a partitioning of the error in a DFT calculation,
\begin{align}
    \Delta E & =\tilde{E}[\tilde{\rho}] - E[\rho] 
    = (\tilde{E}[\tilde{\rho}] -\tilde{E}[\rho]) + (\tilde{E}[\rho]- E[\rho])\\
    &= \Delta E_D +\Delta E_F 
    \label{eq:errorbreakdown}
\end{align}
where $\tilde{E}$ is a given approximate density functional, $\tilde{\rho}$ is the self-consistent density from that functional, and $E$ and $\rho$ are the exact quantities. This definition neatly separates density driven errors for a given functional, $\Delta E_D = \tilde{E}[\tilde{\rho}] -\tilde{E}[\rho]$, from functional error obtained with the exact density, $\Delta E_F=\tilde{E}[\rho]- E[\rho] $.

Employing \cref{eq:errorbreakdown}, a framework was constructed for when the application HF-DFT is appropriate, called density corrected DFT or DC(HF)-DFT\cite{kim2013understanding,kim2014ions,kim2015improved,sim2018quantifying,vuckovic2019density,sim2022improving}. When the following two criteria are satisfied, HF-DFT is used; when at least one criteria is not satisfied, the usual self-consistent DFT is used:
\begin{enumerate}
    \item Is the problem density-sensitive? Density sensitivity is defined as, $S = \tilde{E}[\rho^{LDA}]-\tilde{E}[\rho^{HF}] >2$ kcal/mol where $\tilde{E}$ denotes a chosen density functional approximation.
    \item Is the Hartree-Fock density likely to be reliable? This is answered by requiring that the wavefunction is not strongly spin contaminated: $| \delta \langle S^2_\text{HF} \rangle | < 10\%$.
 \end{enumerate}
The first criterion enforces that HF-DFT is only used when a particular molecule or complex has a large density driven error. The second criterion enforces that HF-DFT is only used in cases where HF is expected to behave well - i.e. when the HF density does not contain an artificial mixing of spin states. The use of this acceptance criteria ensures that DC(HF)-DFT is only used when needed to rectify a significant SIE, and when the HF density is believed to be qualitatively correct. In practice, DC(HF)-DFT has been very successful across a range of problems.\cite{kim2018halogen,dasgupta2021elevating,palos2022assessing,rana2022detection,rana2022correcting,morgante2023density}

The success of HF-DFT has traditionally been attributed to a reduction in density-driven error, however recently Kaplan et. al. found that HF-DFT instead succeeds due to a cancellation of error between the delocalizing functional-driven error and overlocalizing HF density-driven error in several cases including barrier heights\cite{kaplan2023understanding, kanungo2023unconventional}. They found that using a density from a density functional approximation expected to be qualitatively correct for systems of interest in fact led to degraded results relative to HF-DFT due to elimination of the favorable cancellation of errors. While HF-DFT can lead to greatly improved accuracy, these examples suggest it may be for the ``wrong'' reasons. Therefore it appears one should be cautious in drawing conclusions about functional and density error from the results of HF-DFT.

To accurately analyze the breakdown of error between functional and density driven error, one needs an accurate yet affordable proxy for the exact density. As already discussed, HF is considered to yield fairly poor densities due to its neglect of electron correlation, so perhaps its recently reported failure to eliminate density error is unsurprising. However, typical sources of such high quality densities such as orbital optimized coupled cluster theory\cite{sherrill1998energies} are too expensive for general use. Previously, well-behaved density functionals were used to yield ``correct'' densities, but the determination of what may be considered a ``well-behaved'' density functional for a specific use case is a complex problem itself. One therefore needs a source of high-quality densities that is of similar cost to obtain as DFT and does not exhibit the same sensitivity as DFT. 

One such source of improved densities is regularized orbital optimized second order M\o ller-Plesset theory, $\kappa$-OOMP2\cite{lee2018regularized}; by including MP2 correlation energy in its orbital optimization, $\kappa$-OOMP2 approximates Brueckner orbitals, leading to better densities and reducing spin contamination. $\kappa$-OOMP2 has been shown to yield more accurate results for thermochemistry and noncovalent interactions than either HF or conventional MP2\cite{rettig2022revisiting}. Furthermore, the use of $\kappa$-OOMP2 densities in place of HF densities was already seen to significantly improve correlated wavefunction theory\cite{bertels2019third, rettig2020third, loipersberger2021exploring, bertels2021polishing}. When using a regularizer value of $\kappa=1.1$, $\kappa$-OOMP2 removes divergent energy corrections due to the application of perturbation theory on systems with small orbital energy gaps, broadening the number of systems to which OOMP2 (or MP2)\cite{shee2021regularized} may be applied. We therefore expect $\kappa$-OOMP2 to yield an improved approximation to the exact density across many use cases relative to HF. In particular, $\kappa$-OOMP2 ameliorates the erroneous spin-symmetry breaking and overlocalizing of HF. Additionally, $\kappa$-OOMP2 may be performed with only a moderate increase in computation time - the per iteration cost is similar to that of a double-hybrid density functional.\cite{martin2020empirical} We therefore will use $\kappa$-OOMP2 densities within the DC-DFT framework in this work, to better remove density-driven errors.

This approach enables us to perform a systematic investigation to reveal the extent to which fortuitous error cancellation is responsible for the success of HF-DFT. We approach this task by using $\kappa$-OOMP2-DFT (i.e. DFT single point calculations with the $\kappa$-OOMP2 density) with a variety of density functional approximations across barrier heights, halogen noncovalent interactions,  and more. We seek to both further elucidate favorable error cancellation in HF-DFT and its breakdown, and explore the prospects for non-selfconsistent DFT energy evaluation using densities that are better than HF.

\section*{Computational Details}

Single point SCF calculations were performed to obtain all densities. Three densities will be employed to evaluate chemically relevant energy differences using a given density functional. First is the density optimized self-consistently with that functional (i.e. its native density). Second is the Hartree-Fock (HF) density. Third is the density extracted from the occupied orbitals of $\kappa$-OOMP2 calculation. For the latter two cases (HF-DFT and $\kappa$-OOMP2-DFT), the imported density was used to perform a single DFT Fock build and energy evaluation to obtain a non-self-consistent energy. 

Note that the density taken from the $\kappa$-OOMP2 calculations corresponds to the optimized reference determinant (i.e. not including the correlation correction, so no fractional occupation numbers occur). 
All $\kappa$-OOMP2 calculations are performed using the resolution of the identity approximation with the corresponding auxiliary basis sets\cite{weigend2002a}. Following guidelines recommended in previous work,\cite{shee2021regularized, rettig2022revisiting} we used $\kappa=1.1$ in all $\kappa$-OOMP2 calculations.

To assess the improvement that is possible by replacing a native optimized density by either the HF or $\kappa$-OOMP2 density, we selected 7 representative density functionals that span the first 4 rungs of Perdew's classification ladder. From rung 1, we selected the SPW92 version\cite{SPW92functional} of the local spin density approximation. From rung 2, we selected the PBE functional\cite{PBEfunctional}, probably the most widely used generalized gradient approximation (GGA). From rung 3, we chose the recent r$^2$SCAN functional\cite{r2SCANfunctional}, which represents a significant improvement over the original SCAN meta-GGA.\cite{SCANfunctional} From rung 4, we selected two global hybrids, B3LYP and PBE0,\cite{B3LYPfunctional, PBE0functional} a range-separated hybrid GGA, $\omega$B97X-V\cite{wB97X-Vfunctional} and a range-separated hybrid meta-GGA, $\omega$B97M-V.\cite{wB97M-Vfunctional}
For the Hait Dipole set, geometries from the 2018 paper were used, with reference values calculated with CCSD(T) extrapolated to the complete basis set limit.\cite{hait2018accurate}. We calculated the dipoles via finite difference method using aug-cc-pVTZ basis set. For the BH76 and SIE4x4 datasets, GMTKN55 database geometries were used with reference barrier heights obtained via the W2-F12 method\cite{GMTKN55dataset}. We perform our approximate DFT calculations using the aug-cc-pVQZ basis set \cite{kendall1992a}. For the Bauz\'a dataset, geometries were used with reference binding energies obtained via  CCSD(T) with 50\% counterpoise correction extrapolated to the complete basis set limit \cite{Halogenbonding, bauzaset}. We performed our approximate DFT calculations using the aug-cc-pVQZ basis set with counterpoise correction.

All calculations were performed using the Q-Chem software package\cite{epifanovsky2021software}.

\section*{Results}
\subsection*{Dipole Moments}
The multipole expansion is a convenient way to express the electrostatic potential and the electric field generated by the charges inside a molecule. For neutral molecules the leading term is the electric dipole moment, which can be understood classically as the sum of charges $q_i$ weighted by their positions $r_i$, or the corresponding quantum expectation value, giving:
\begin{equation}
    \vec{\mu}= \vec{\mu}_e + \vec{\mu}_n = \sum_A^{N_{\textrm{nuc}}} Z_A \vec{R}_A - \mathrm{Tr}\left[ \mathbf{p} \vec{\mathbf{r}} \right]
\end{equation}
Here $\mathbf{p}$ is the one-particle density matrix, and $\vec{\mathbf{r}}$ is the position operator in the atomic orbital basis. For neutral molecules, the dipole moment is independent of origin, making it the simplest measure of the charge distribution: its first moment which characterizes polarity along the $x, y, z$ axes. Higher moments can likewise be defined.\cite{hait2021too}  on through which we can compare the quality of the densities obtained via a given electronic structure method.

When considering the dipoles produced by DC-DFT methods, it might be assumed that the dipole moment would be shared with that of the reference density. However, this is incorrect because the Hellmann-Feynman theorem is not applicable to non-variational methods such as DC-DFT. Instead, the dipole moment must be defined and evaluated as the first order response of the molecular energy to an electric field $\vec{\mathcal{E}}$. 
\begin{equation}
     \vec{\mu}_e= \frac{d E}{d \vec{\mathcal{E}}} = \frac{\partial E}{\partial \mathbf{h}} \frac{\partial \mathbf{h}}{\partial \vec{\mathcal{E}}} + \frac{\partial E}{\partial \boldsymbol{\theta}} \frac{\partial \boldsymbol{\theta}}{\partial \vec{\mathcal{E}}}
    \label{eq:responsedipole}
\end{equation}
Here the symbol ``d" reminds us that we take the total derivative, and if there are non-optimized rotation parameters (as in DC-DFT), $\boldsymbol{\theta}$, then two separate partial derivative contributions arise, as given above. The latter contains their electric field dependence contributes via the second (non-Hellmann-Feynman) contribution. The first (where $\mathbf{h}$ is the one-body hamiltonian) yields the expectation value contribution, $- \mathrm{Tr}\left[ \mathbf{p} \vec{\mathbf{r}} \right]$.  

We investigate a dipole set of 152 molecules\cite{hait2018accurate}.  In  \cref{tab:Dipoles}, the mean absolute regularized error (MAE) is reported, as defined from the regularized errors 
\begin{equation}
    r\text{Error}=\frac{\mu-\mu_{ref}}{\text{max}(\mu_{ref},1D)} \times 100\%
    \label{eq:reg_error}
\end{equation}
The MAEs are separated between non spin polarized (NSP) that used  restricted orbitals, and spin polarized (SP) which used unrestricted orbitals. The dipoles are computed using \cref{eq:responsedipole} via finite differences.\cite{hait2018accurate} The $\kappa$-OOMP2 dipoles have two values, with $\mu_0$ using the reference determinant from $\kappa$-OOMP2 to compute the DFT energy, and $\mu$ using the rigorous 1 particle density matrix which includes the non-integer occupations associated with double substitutions of occupied orbitals by virtuals used in $\kappa$-OOMP2.

\begin{table}[htbp]
   \centering
   \begin{tabular}{|cc|cccc |}
   \hline
                  & &   DFT & HF & $\kappa$-OOMP2 & $\kappa$-OOMP2  \\ 
                  & & & & $\mu_0$ & $\mu$\\
    \hline
SPW92	&	NSP	&	6.21	&	4.45	&	5.74	&	5.82		\\	
	&	SP	&	11.13	&	18.69	&	10.97	&	9.40		\\	
	&	All	&	8.54	&	11.20	&	8.22	&	7.51		\\	\hline
PBE	&	NSP	&	6.06	&	3.40	&	5.34	&	5.73		\\	
	&	SP	&	9.48	&	14.85	&	9.77	&	8.81		\\	
	&	All	&	7.68	&	8.82	&	7.44	&	7.19		\\	\hline
r$^2$SCAN	&	NSP	&	5.01	&	3.62	&	4.72	&	5.28		\\	
	&	SP	&	7.24	&	13.22	&	7.30	&	10.25		\\	
	&	All	&	6.07	&	8.17	&	5.94	&	7.63		\\	\hline
PBE0	&	NSP	&	3.32	&	2.95	&	3.17	&	4.08		\\	
	&	SP	&	4.68	&	13.35	&	4.65	&	8.04		\\	
	&	All	&	3.96	&	7.88	&	3.87	&	5.96		\\	\hline
B3LYP	&	NSP	&	2.79	&	2.58	&	2.55	&	3.20		\\	
	&	SP	&	5.16	&	15.45	&	4.65	&	7.09		\\	
	&	All	&	3.91	&	8.68	&	3.54	&	5.04		\\	\hline
$\omega$B97X-V	&	NSP	&	3.78	&	3.68	&	3.73	&	5.16		\\	
	&	SP	&	5.30	&	17.47	&	6.13	&	10.27		\\	
	&	All	&	4.50	&	10.21	&	4.87	&	7.58		\\	\hline
$\omega$B97M-V	&	NSP	&	2.78	&	2.67	&	2.75	&	3.46		\\	
	&	SP	&	5.30	&	19.53	&	6.64	&	9.70		\\	
	&	All	&	3.97	&	10.66	&	4.59	&	6.42		\\	\hline
Self-	&	NSP	&	-	&	10.50	&	4.02	&	3.22		\\	
Consistent	&	SP	&	-	&	18.38	& 5.92		&	5.36		\\	
	&	All	&	-	&	14.23	&	4.92	&	4.23		\\	\hline

\end{tabular}
    \caption{Regularized MAE (as defined in \cref{eq:reg_error}) for 152 dipoles.\cite{hait2018accurate} For each functional, the columns report MAE evaluated using that functional with (i) DFT: the self-consistent DFT density, (ii) HF: use of the HF density for HF-DFT energies, (iii) $\kappa$-OOMP2 $\mu_0$: use of the $\kappa$-OOMP2 reference determinant (iv) $\kappa$-OOMP2 $\mu$: use of the full $\kappa$-OOMP2 1-particle density matrix. For comparison, the row labeled self-consistent reports MAE in dipoles for the HF density, the $\kappa$-OOMP2 reference determinant, and the $\kappa$-OOMP2 1-particle density matrix. MAE is reported for the spin polarized (SP) and not spin polarized (NSP) cases separately, as well for all cases.}
    \label{tab:Dipoles}
\end{table}

Beginning with the self-consistent row in \cref{tab:Dipoles}, we see that HF provides a qualitatively correct, yet quantitatively poor description of the dipole moment with 11\% MAE for the NSP class, and a significantly worse 18\% MAE for the SP cases. By contrast, $\kappa$-OOMP2 provides a significantly improved description of the dipoles, reducing the MAE by over a factor of 3 to 3\% MAE for the NSP cases, and 5\% MAE for the SP cases. The dipole moment  $\mu_0$ from the $\kappa$-OOMP2 reference determinant follows the same trends, albeit slightly degraded due to its neglect of fractional occupations. These results provide quantitative evidence that employing the $\kappa$-OOMP2 reference determinant for the later tests is, in fact, employing a significantly more accurate electron density than using HF.

Comparison across density functionals evaluated at self-consistent, HF, and $\kappa$-OOMP2 reference densities show a clear trend. For NSP cases, HF densities provide improved dipole moments (relative to use of the native density) across all levels of functionals. The improvement is largest for rungs 1, 2, and 3. The results also exhibit MAE statistics that are much better than the HF density itself for the NSP class! This is not a contradiction: it means that the second (non-Hellmann-Feynman) term in \cref{eq:responsedipole} greatly improves the result over use of just the first term (which of course returns the HF statistics, unmodified). This provides very direct evidence for the ``importance of being non-selfconsistent''\cite{wasserman2017importance} when it comes to using HF-DFT for dipole moments for normal closed shell molecules!

What happens when we use a better density? For functionals on rungs 1,2, and 3, use of the $\kappa$-OOMP2 density (i.e. $\kappa$-OOMP2-DFT) from either the reference determinant or the full 1PDM significantly degrades the very good NSP HF-DFT results. Since the Hellmann-Feynman term is \textit{better} than HF-DFT (it yields the native $\kappa$-OOMP2 dipole as opposed to the native HF dipole), this reveals that the non-Hellmann-Feynman term \textit{degrades} that improvement. In other words, the opposite issue that occured in HF-DFT! Use of a better density degrades the NSP dipole moments evaluated non-self-consistently via low rung functionals. Translating this into the terminology of DC-DFT (\cref{eq:errorbreakdown}), we can say that the success of HF-DFT reflects error cancellation in these low-rung functionals between density-driven errors and functional error.

When looking at the more difficult SP cases, the HF density worsens the dipole, regardless of level of density functional approximation. Given that a third of the SP dipoles in this set are spin contaminated, this reinforces the need for DC-DFT criterion 2 at the HF level. The $\kappa$-OOMP2 1PDM density provides improvements over HF in the SP cases, but remain around 10\% MAE. Significant improvement can be observed from the reference determinant density from  $\kappa$-OOMP2 where it reduces errors from the HF reference by roughly a half and from the 1PDM density by a third for most functionals. The accuracy of the reference determinant dipole in both self-consistent and DC-DFT calculations suggests that it is an appropriate proxy for the exact density to be used in DC-DFT error analysis.

\subsection*{Barrier Heights}
We investigated the BH76 dataset, a dataset of 76 reaction barrier heights involving main group elements. DFT, especially with local functionals, is known to underestimate barrier heights\cite{janesko2008hartree,verma2012increasing}  due to density driven errors associated with SIE. Recent work\cite{DCDFTQ&A} has noted that HF-DFT with the PBE functional yielded a reduction in error from 9.0 (self-consistent PBE) to 4.4 kcal/mol. We performed analogous calculations, using the optimized DFT density, the HF density and the $\kappa$-OOMP2 density for a set of 7 representative density functionals. As $\kappa$-OOMP2 is known to yield significantly better densities than HF, it is expected that $\kappa$-OOMP2-DFT will perform at least as well as HF-DFT if the poor performance of DFT is due to a density driven error. 

\begin{figure}[h]
    \centering
    \includegraphics[width=0.7\textwidth]{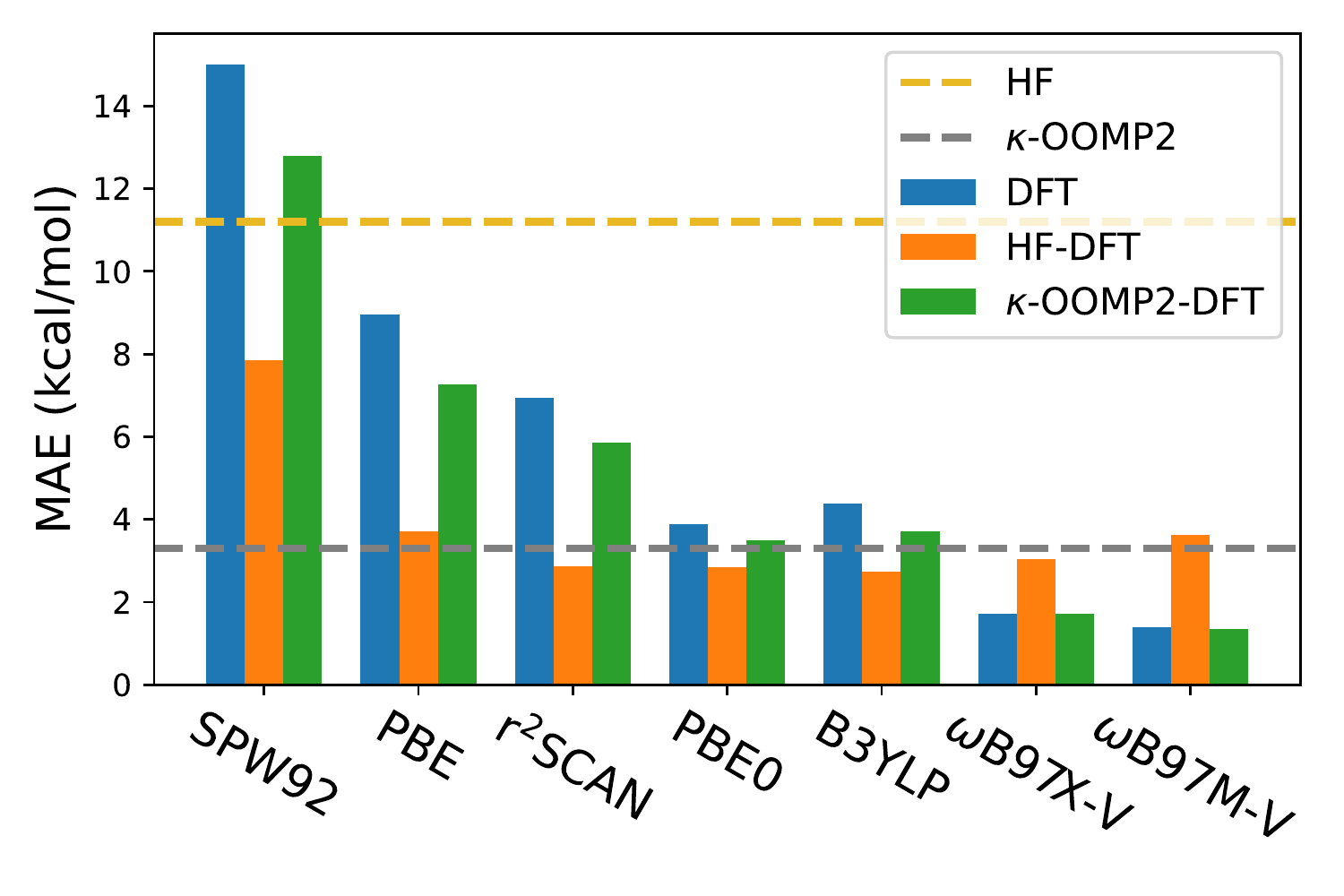}
    \caption{BH76 mean absolute errors for various density functionals, each evaluated using 3 different densities (from left to right, the self-consistently optimized density, the HF density, and the $\kappa$-OOMP2 density. The upper and lower dashed lines correspond to the MAE for HF and $\kappa$-OOMP2, for comparison.}
    \label{fig:bh76data}
\end{figure}

The statistics obtained from our calculations are summarized in graphical form in \cref{fig:bh76data} and in  tabular form in  \cref{tab:BH76}.
For PBE, $\kappa$-OOMP2-DFT does improve upon DFT, reducing the MAE from 9.0 kcal/mol to 7.3 kcal/mol, but not nearly to the extent of HF-DFT, which achieves an MAE of 3.7. The better density therefore yields nearly two times larger error! Very similar results are seen with $r^2$SCAN, and SPW92 is also qualitatively the same. For these rung 1, 2 and 3 functionals, the results strongly reinforce the conclusion obtained recently by Kaplan and co-workers\cite{Perdew2023Barriers} that the success of DC-DFT using the HF density arises from a fortunate cancellation of errors. The overly localized HF density evidently yields a reasonable energy when using a functional that erroneously lowers the energy of delocalized solutions. 

\begin{figure}
    \includegraphics[width=0.7\textwidth]{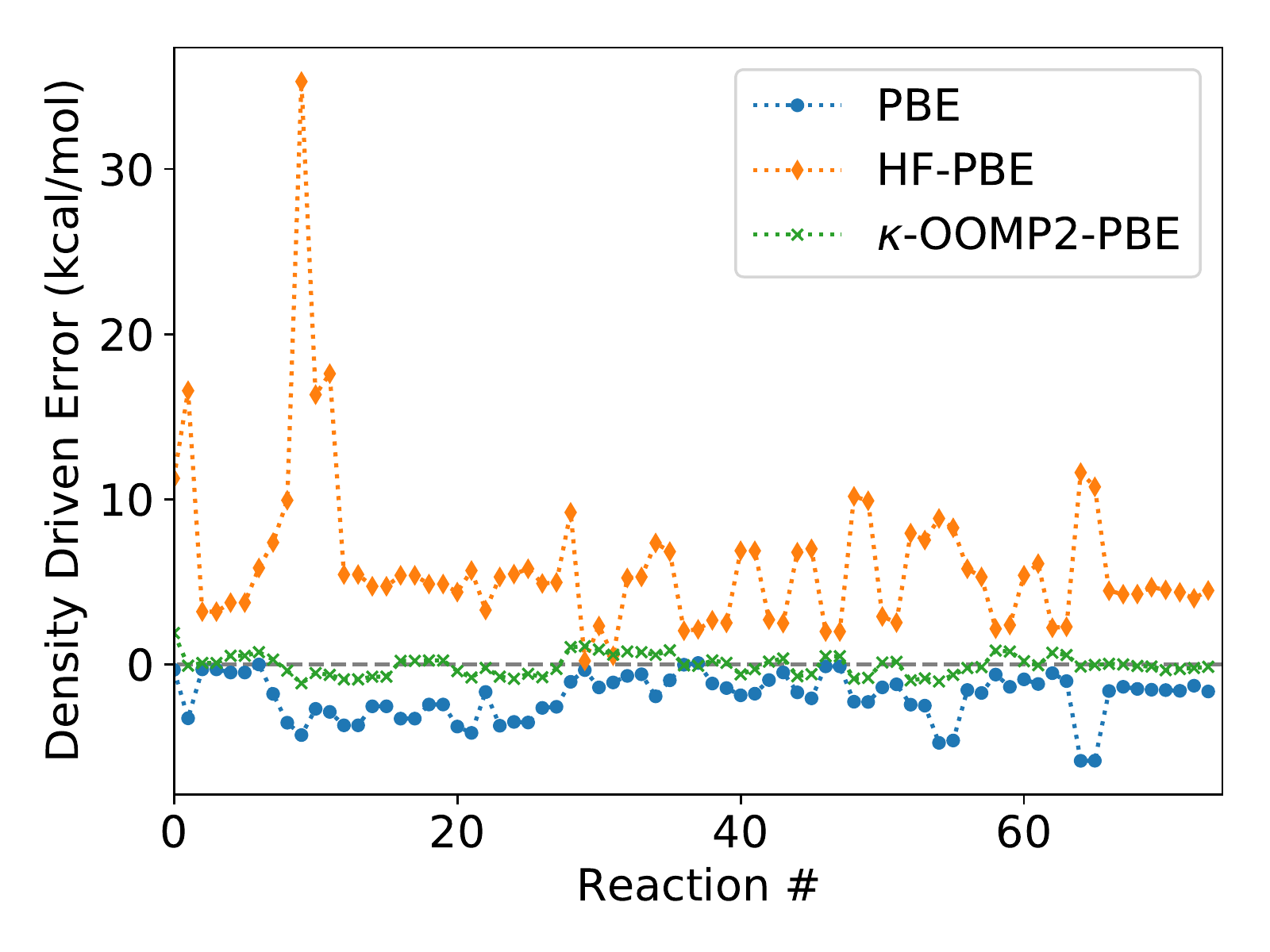}
    \caption{Density-driven error with respect to OD reference orbitals for each of the 76 barriers contained in the BH76 dataset, measured with the PBE functional. The data shows that larger errors arise from using the HF density than the PBE density, while use of the $\kappa$-OOMP2 reference orbitals yields very small errors.}
    \label{fig:od_density_error}
\end{figure}

In order to quantify the density-driven and functional-driven error, we use $\kappa$-OOMP2 as a proxy for the exact density. To further justify this choice, we quantify the accuracy of the underlying $\kappa$-OOMP2 densities. For the BH76 dataset, we additionally utilized one of the highest quality source of orbitals commonly available - optimized coupled cluster doubles\cite{sherrill1998energies} (OD) in a smaller TZ basis. By assuming that the OD density is the reference density, we can compute the density driven and functional driven error according to \cref{eq:errorbreakdown}. We are specifically interested in the density driven error introduced from the $\kappa$-OOMP2 density. In \cref{fig:od_density_error}, we show the density driven error for the PBE functional using PBE, HF, and $\kappa$-OOMP2 densities, relative to the reference OD densities for the reactions in the BH76 dataset. $\kappa$-OOMP2 has a quite small density driven error - less than 1 kcal/mol in nearly all cases - and nearly no systematic bias. PBE density leads to a systematic underestimation while HF leads to an even greater systematic overestimation. We therefore see that $\kappa$-OOMP2 yields a satisfactory approximation to the reference density, which further justifies our use of $\kappa$-OOMP2 as the reference density for the remainder of this study.

\begin{figure}[ht!]
  \centering
  \begin{subfigure}[b]{0.6\linewidth}
    \includegraphics[width=\linewidth]{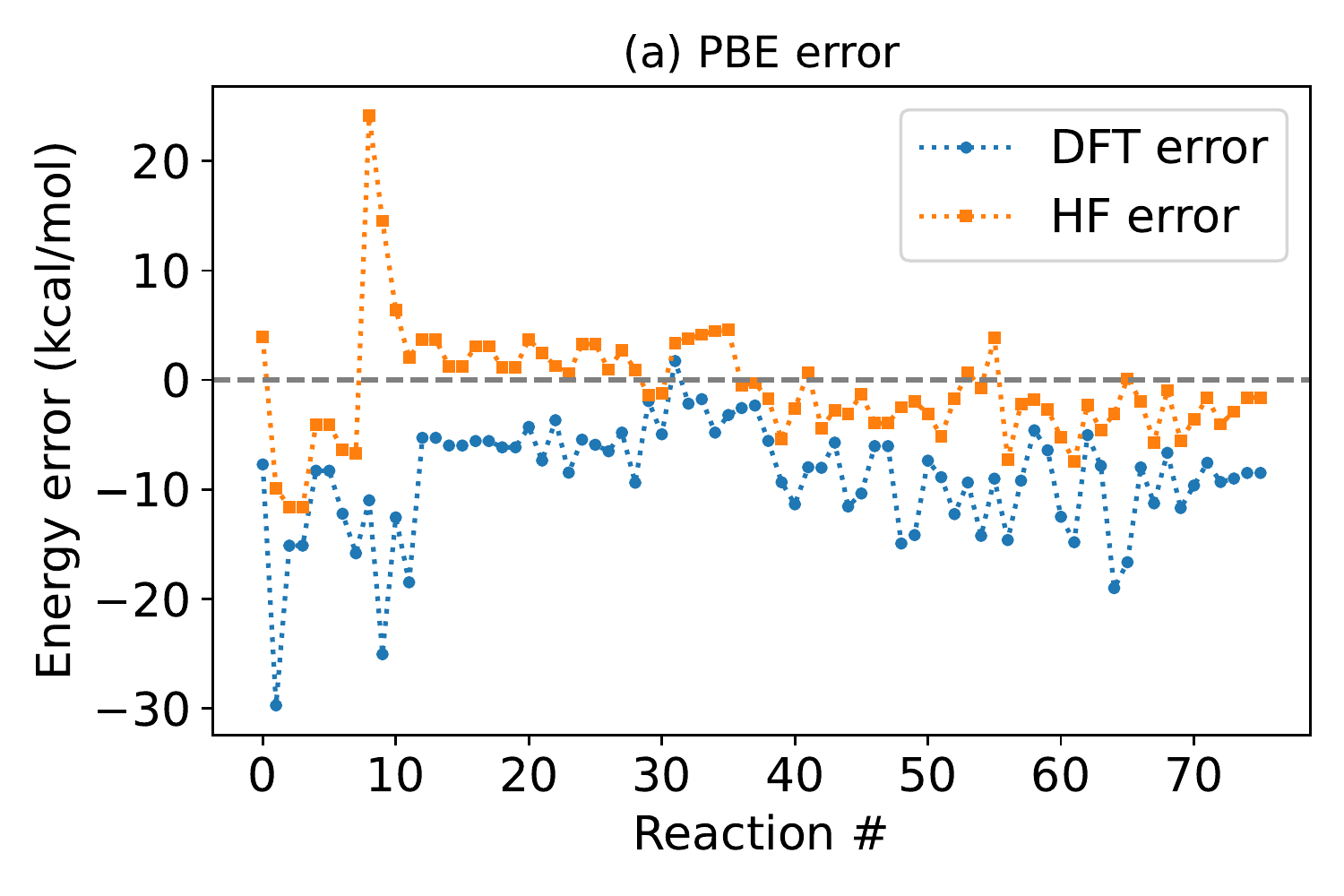}
    
  \end{subfigure}
  \begin{subfigure}[b]{0.6\linewidth}
    \includegraphics[width=\linewidth]{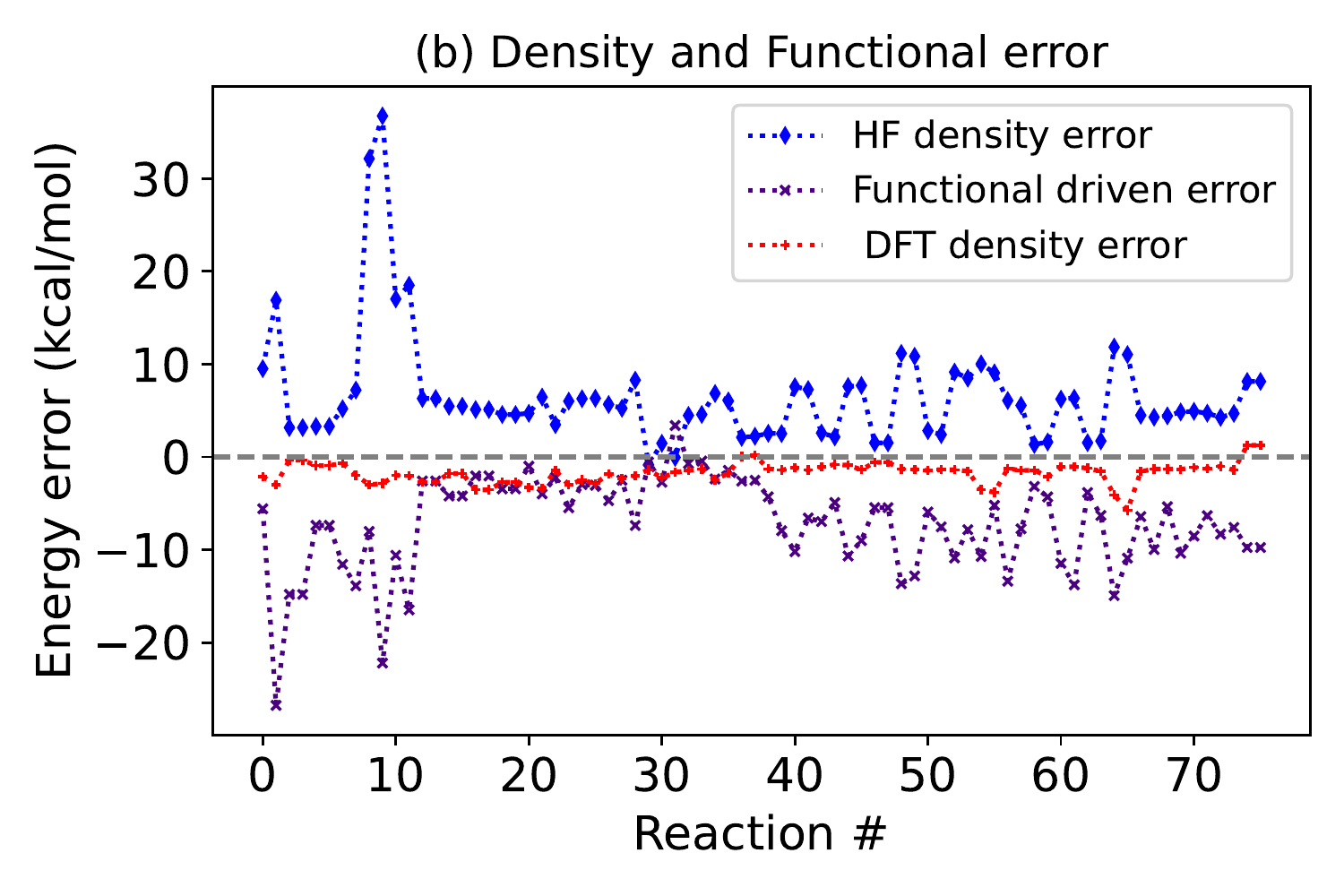}
  \end{subfigure}

\caption{Decomposition of error components PBE functional evaluated at a given reference density in BH76. (a) Plot comparing self-consistent PBE and HF-PBE error for each reaction. The legend denotes which reference density was used. (b) Plot of functional driven error of PBE, density driven error for the functional evaluated self-consistently, and density driven error for the functional evaluated at the HF density.}
\label{fig:PBEerrordecomp}
\end{figure}

Still focusing only on the PBE functional, we show the DFT and HF-DFT error for each individual datapoint within the BH76 dataset in \cref{fig:PBEerrordecomp}a, while \cref{fig:PBEerrordecomp}b shows the breakdown of that error according to \cref{eq:errorbreakdown} (using the $\kappa$-OOMP2 density as a proxy for the true density). The figure clearly shows that rather than ameliorating the density error, HF-DFT actually increases it! In fact, this is actually quite consistent with the relatively poor quality of HF densities that we reviewed in the Introduction. It is evident that the improved performance of HF-DFT on BH76 with the PBE functional (and SPW92 and r$^2$SCAN) may be attributed to a fortunate cancellation of error between the density-driven and functional-driven error.

\begin{table}
    \centering
    \begin{tabular}{|c|ccccc|}
    \hline
                   &   DFT & HF & $\kappa$-OOMP2 & DC(HF) & DC($\kappa$-OOMP2)\\ 
    \hline
     SPW92    & 15.00	&7.85	&12.80&	9.19&	13.23\\
     PBE & 8.95 &	3.72	&7.27& 4.42&	7.59\\
     r$^2$SCAN & 6.94	&2.87	&5.85&	3.63&	6.07\\
     PBE0 & 3.88	&2.84	&3.49&	2.98&	3.65\\
     B3LYP & 4.39	&2.73	&3.72	&2.69&	3.90\\
     $\omega$B97X-V & 1.71	&3.04&	1.71&		1.56&	1.70\\
     $\omega$B97M-V & 1.40	&3.63	&1.35&	2.43&	1.41\\
     Self-Consistent & -& 11.29&3.81 & -& - \\
    \hline
    \end{tabular}
    \caption{Mean absolute error for BH76 set reported in kcal/mol. The rows represent density functional approximations used to evaluate the reference density of a given column. }
    \label{tab:BH76}
\end{table}

So far, we have concentrated on DC-DFT for the PBE functional. However, \cref{fig:bh76data} also shows the effect of different densities on the MAE for BH76 for a range of other functionals. It is apparent that the helpful effects of using the HF density are reduced as one climbs higher on Jacob's ladder. Significant improvements are seen for low rung functionals, but for range separated functionals where the SIE is reduced, the density-driven error and functional-driven error are both small so the increase in density driven error from using the HF density does not lead to error cancellation - it instead leads to significantly worse performance. 

The $\kappa$-OOMP2 reference density makes modest improvements over self-consistent DFT for all functionals, converging to the DFT results near the top of Jacob's ladder. As we are approximating the exact density as the $\kappa$-OOMP2 density, $\kappa$-OOMP2-DFT errors are expected to be dominated by functional-driven error. It is therefore noteworthy that $\kappa$-OOMP2-DFT performs significantly worse than HF-DFT for low-rung functionals as the elimination of density-driven error also eliminates the fortuitous cancellation of error.

On the right side of \cref{tab:BH76}, we include the density corrected versions of each method, where the ``correct" density is only used if it is believed to be suitable. For the HF reference density, we see that the DC method performs worse on average than HF-DFT. 
Evidently, the reactions we expect to not benefit from using HF density (either due to density insensitivity or qualitative breakdown of HF), still yield significant improvements over self-consistent DFT - further suggesting a fortuitous cancellation of error. 
Comparing the methods with $\kappa$-OOMP2 as the reference density, we see an order of magnitude smaller shifts between their errors. This suggests that in the cases where density driven error is not strong, $\kappa$-OOMP2 still captures the correct physics.

We may probe further into the discrepancies between HF-DFT and DC(HF)-DFT, to know the best time, place, and occasion to use the flavors of DC-DFT.
Within BH76, there is a subset of 12 reactions that exhibit spin contamination when using HF densities. Following the guidelines for DC-DFT, we do not use the HF density for these cases, leading to the differing performance of DC(HF)-DFT and HF-DFT. 
Despite the known poor quality of the HF density for these cases, HF-DFT is shown to still perform well for these cases when using pure functionals, while the $\kappa$-OOMP2 density, which correctly removes spin contamination, does not provide significant improvement over self-consistent DFT. HF-DFT therefore benefits from using qualitatively incorrect spin contaminated densities, leading to its superior performance to DC(HF)-DFT. 

\begin{figure}[ht!]
    \centering
\includegraphics[width=0.7\textwidth]{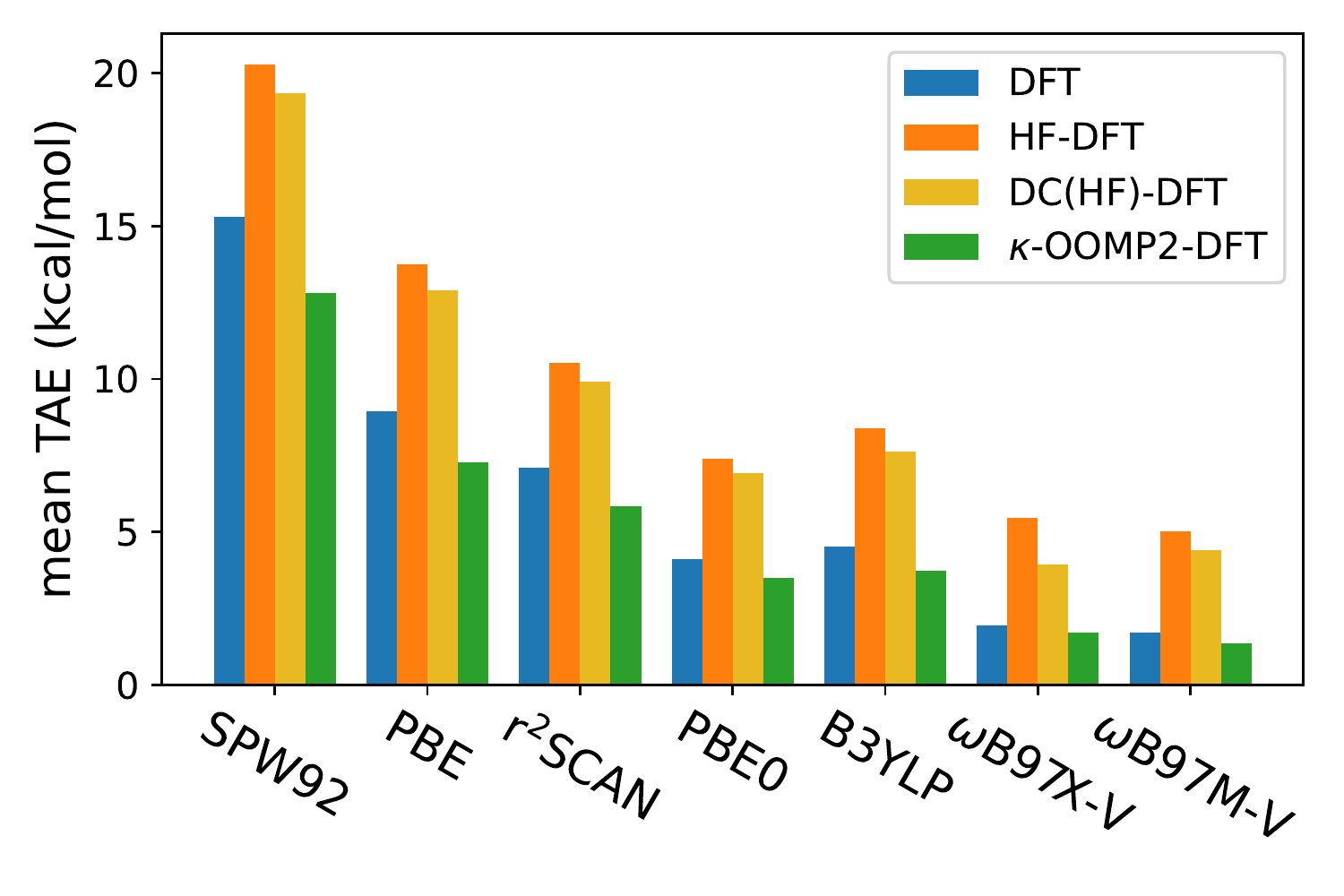}
    \caption{Chart displaying $\Delta E_{TA}$ in kcal/mol for BH76, varying both functional and reference density.  }
 \label{fig:bh76totalerror}
\end{figure}

To determine whether a given functional obtains correct energies for the ``correct'' reasons, the fortuitous cancellation of error may be eliminated by introducing an expression for the total absolute (TA) error in a DFT calculation.
\begin{equation}
    \Delta E_{TA} = |\Delta E_D |+|\Delta E_F|
    \label{eq:eta}
\end{equation}
We plot $\Delta E_{\text{TA}}$ for BH76 in figure \cref{fig:bh76totalerror}. This plot shows behavior more consistent with the climb up Jacob's ladder to reduce errors than \cref{fig:bh76data}. We see that using the HF density increases $\Delta E_{\text{TA}}$, as the density actually becomes \textit{worse}, increasing $\Delta E_{\text{D}}$. As we use $\kappa$-OOMP2 as our reference ``exact" density, the $\kappa$-OOMP2-DFT $\Delta E_{\text{TA}}$ is equivalent to $\Delta E_{\text{F}}$. As expected, we see that the DFT $\Delta E_{\text{TA}}$ converges towards $\Delta E_{\text{F}}$ as you ascend Jacob's ladder - i.e. higher rung functionals yield higher quality densities.

Overall, it seems abundantly clear that HF-DFT with low rung functionals performs very well for BH76 by combining poor functionals with poor densities. This presents a mild philosophical dilemma, as DC(HF)-DFT still outperforms $\kappa$-OOMP2 based methods with less computational cost.

\subsection*{Halogen Bonding}

\begin{figure}[ht!]
    \centering
    \includegraphics[width=0.7\textwidth]{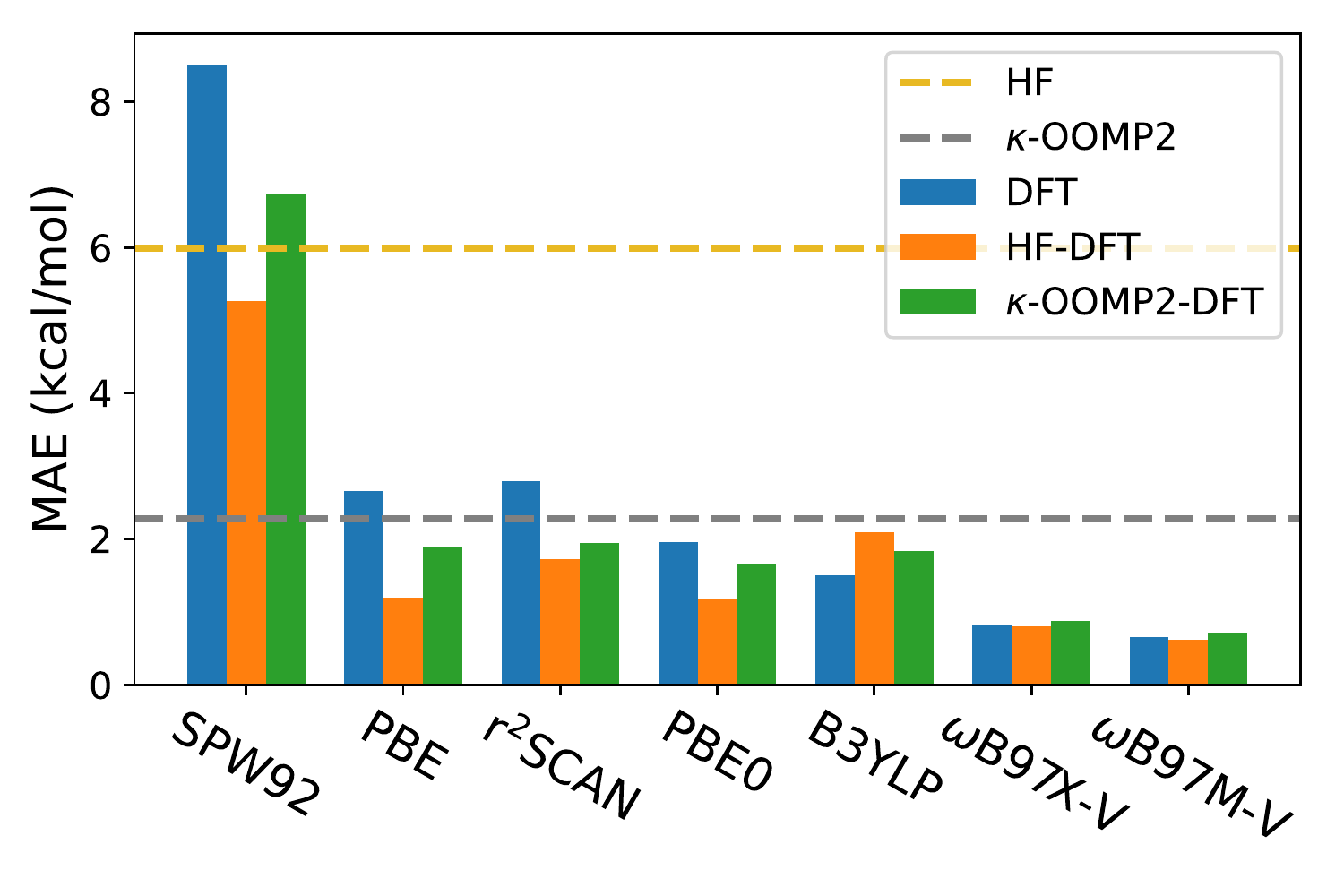}
    \caption{Bauz\'a set mean absolute errors for various density functionals  evaluated at a given density. HF and $\kappa$-OOMP2 are plotted as energy references.}
    \label{fig:bauzadata}
\end{figure}

\begin{figure}[ht!]
    \centering
    \includegraphics[scale=1]{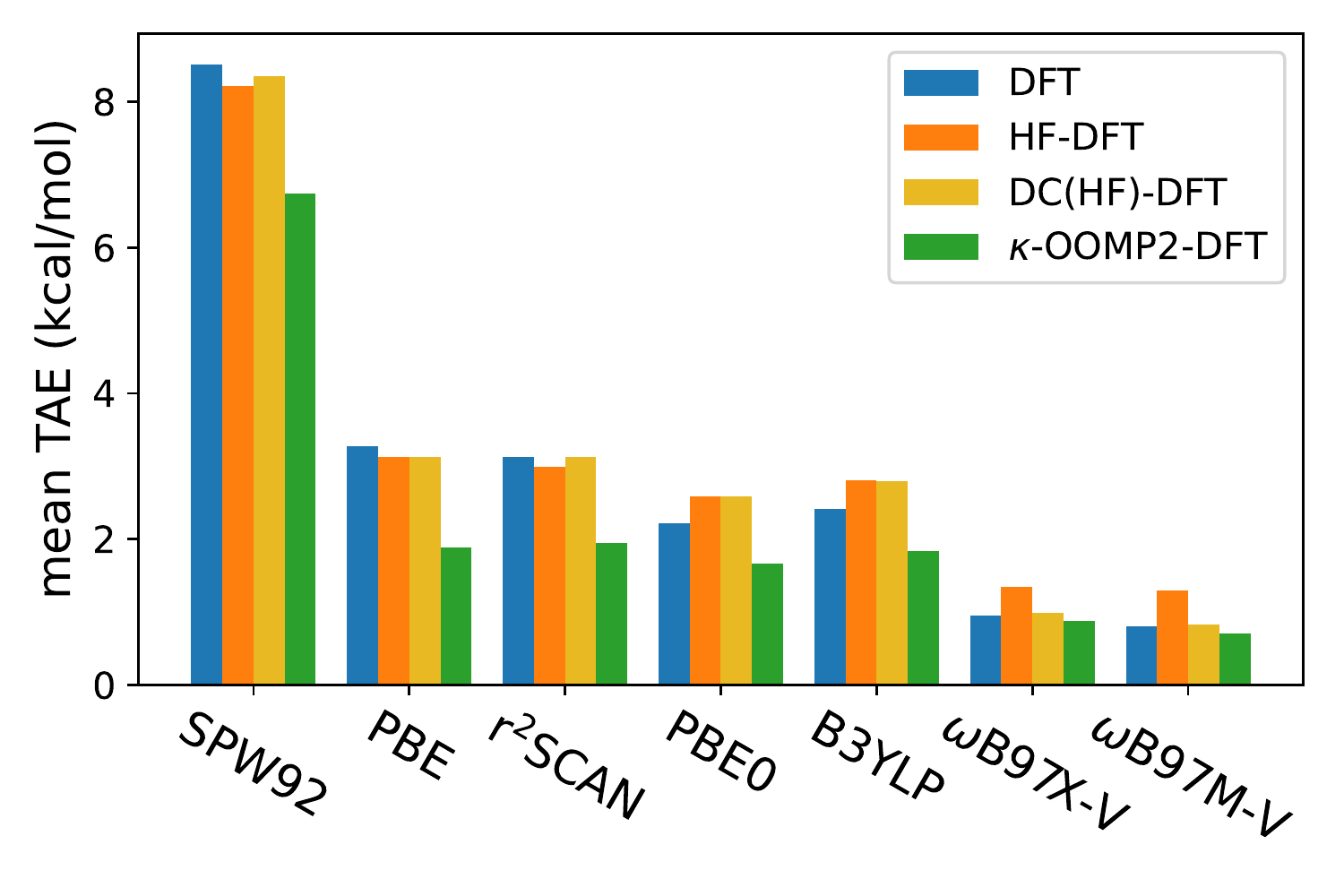}
    \caption{Bauz\'a set total absolute error, $\Delta E_{TA}$, for various density functionals  evaluated at a given density.}
    \label{fig:bauza_totalerror}
\end{figure}

\begin{figure}[ht!]
  \centering
  \begin{subfigure}[b]{0.6\linewidth}
    \includegraphics[width=\linewidth]{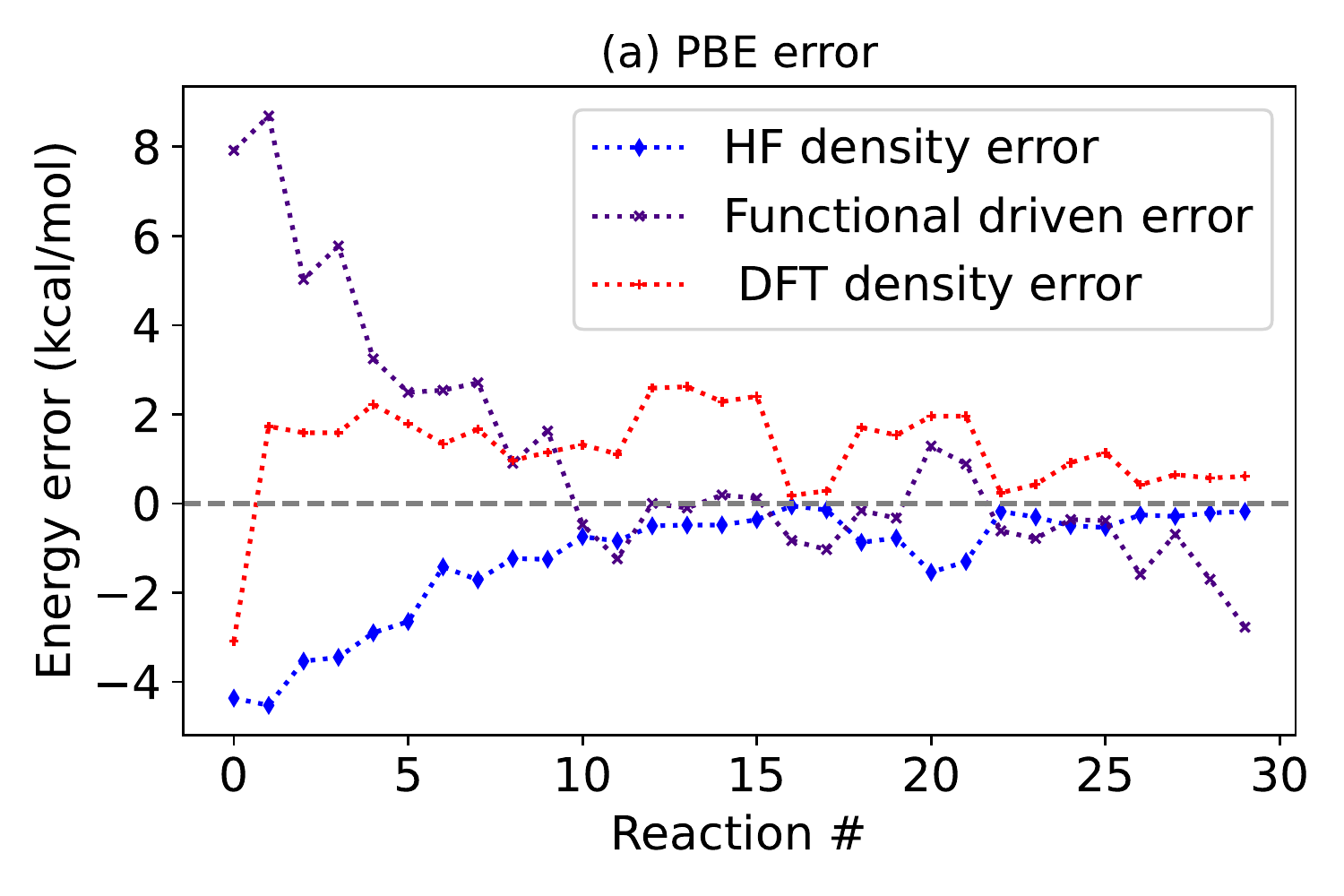}
   
    \label{fig:bauzatotal_pbe_error}
  \end{subfigure}
  \begin{subfigure}[b]{0.6\linewidth}
    \includegraphics[width=\linewidth]{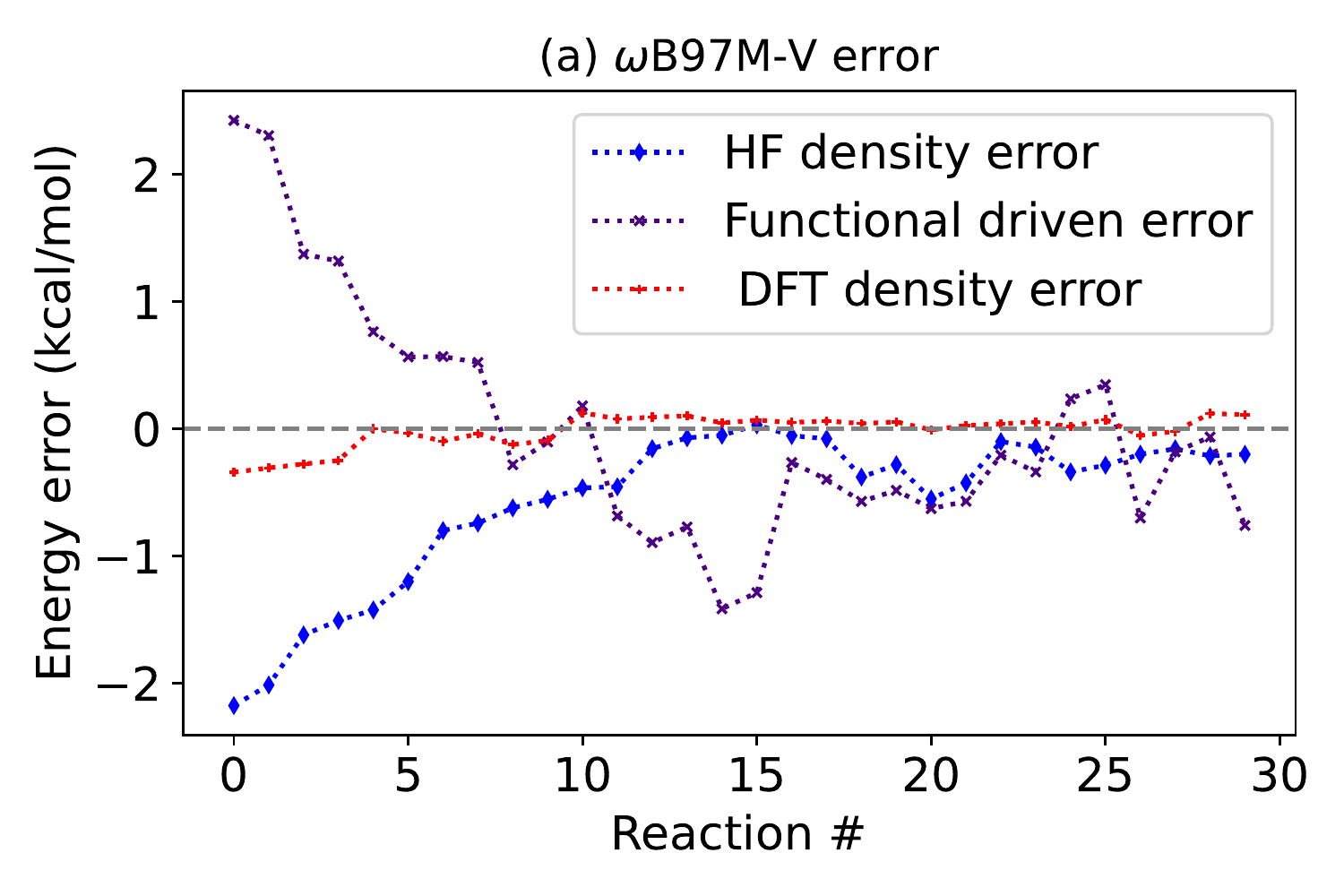}
    \label{fig:bauzadft_densityerror}
  \end{subfigure}
  \\
\caption{Bauz\'a  set error decomposition comparing (a) the case of PBE, where HF-DFT reduces mean total absolute error, and (b) $\omega$B97M-V, where it does not.}
\label{fig:bauzaerrordecomp}
\end{figure}

In light of this very good performance of HF-DFT for barrier heights by cancellation of errors, we sought to investigate the origin of other notable successes of DC-DFT. We next looked at the Bauz\'a set of binding energies for 30 halogen bonded dimers, where this set is selected for the presence of charge transfer in the binding,\cite{thirman2018characterizing} leading to considerable self-interaction error\cite{Halogenbonding}. Mean absolute errors for HF and $\kappa$-OOMP2 based DC-DFT methods are shown in \cref{fig:bauzadata} and  \cref{tab:Bauza}. We see similar trends as for the BH76 set where HF-DFT yields significant reductions in error over DFT while $\kappa$-OOMP2-DFT yields improved errors to a lesser degree. For this dataset however, there is no transition where the HF reference begins to degrade the binding energies over DFT densities; HF-DFT performs nearly identically to self-consistent DFT for range separated hybrids. Additionally, $\kappa$-OOMP2-DFT performs only slightly worse than HF-DFT in most cases, and the degradation from HF-DFT to DC(HF)-DFT is quite minor compared to BH76. This evidence all suggests that HF-DFT does not benefit from nearly as large a degree of error cancellation in this case. Looking at the total absolute error for the Bauz\'a set in \cref{fig:bauza_totalerror} confirms this, with HF and DC(HF) methods having a similar magnitude to DFT. Comparing this result to the MAE's for the set reveals that growing errors in HF-DFT relative to DFT are what allow it to retain similar accuracy to DFT at rung 4 of DFT. Given that the MAE and total absolute error is reduced for LDA and GGA type functionals, it can be said that HF-DFT is providing a better density in these cases. Even so, \cref{fig:bauzaerrordecomp} shows that for cases where HF-DFT performs well and reduces error, it still relies on large over-corrections to achieve these results.

\begin{table}[ht!]
   \centering
   \begin{tabular}{|c|ccccc |}
   \hline
                   &   DFT & HF & $\kappa$-OOMP2 & DC(HF) & DC($\kappa$-OOMP2) \\ 
    \hline
     SPW92  &8.51&	5.26&	6.74		&5.60	&6.88
\\
     PBE & 2.82	&1.20&	1.88	&	0.91&	1.73
 
\\
r$^2$SCAN & 2.80	&1.73&	1.95		&1.93	&2.11

\\
PBE0 & 1.96&	1.18&	1.66&		1.42	&1.87
\\
B3LYP & 1.50	&2.10 &	1.83		&1.17	&1.40
\\
$\omega$B97X-V &0.83&	0.80	&0.88&		0.87	&0.84  
\\
$\omega$B97M-V &0.65&	0.62&	0.71	&	0.71	&0.67\\
Self-Consistent& -&5.99 &2.28 & -& - \\
\hline
\end{tabular}
    \caption{Mean absolute error for Bauz\'a set reported in kcal/mol}
    \label{tab:Bauza}
\end{table}

\subsection*{Bond Breaking}
Lastly, the SIE4x4 subset of GMTKN55\cite{goerigk2017look} was investigated. These  small dimer cations with stretched bonds represent a worst case scenario for self-interaction error leading to very erroneous results for many functionals, as can be anticipated from \cref{fig:h2diss}. For bond breaking problems like these, approximate density functionals often reach incorrect dissociation limits with fractional charges due to SIE. By contrast, HF breaks spin symmetry out of necessity for these problems and provides a qualitatively correct  picture of the density at the dissociated limit (i.e. localized charges) for these systems.

The MAE statistics for various functionals using 3 different densities are shown in \cref{tab:SIE}. Here we see that HF-DFT again significantly improves the poor energetics of DFT. The improvements are also again largest for the lower rung functionals. Despite these improvements over use of the self-consistent DFT density, HF-DFT still yields errors that range between 7 and 13 kcal/mol for all functionals. By contrast, the $\kappa$-OOMP2  reference orbitals provide only modest improvements over self-consistent DFT orbitals, and are significantly poorer than using HF orbitals. We note that $\kappa$-OOMP2 itself provides an MAE of 2.4 kcal/mol, which reminds us that the reference density is quite accurate. 

\begin{table}[h]
    \centering
    \begin{tabular}{|c|ccccc |}
    \hline
	& DFT &	HF	& $\kappa$-OOMP2	& DC(HF)&	DC(kOOMP2)\\
 \hline
SPW92&	27.64	&13.43&	23.41&	13.45&	23.43
\\
PBE &	23.24&	10.62	&18.77&	10.91&	19.06
\\
r$^2$SCAN&	17.97&	9.03	&14.55&	9.35	&14.81
\\
PBE0 & 14.03&	7.69	&11.57	&7.84	&11.73
\\
B3LYP & 24.07&	9.45&	14.64&	18.74&	23.94
\\
$\omega$B97X-V	&11.40	&7.63	&9.99&	7.98	&10.14
\\
$\omega$B97M-V	&10.64	&6.84&	9.10&	7.10	&9.37
\\
Self-Consistent & -& 9.18 &2.43 & -& -\\

    \hline
    \end{tabular}
    \caption{Mean absolute error for SIE4x4 set reported in kcal/mol}
    \label{tab:SIE}
\end{table}

\begin{figure}[!h]
  \centering
  \begin{subfigure}[b]{0.6\linewidth}
    \includegraphics[width=\linewidth]{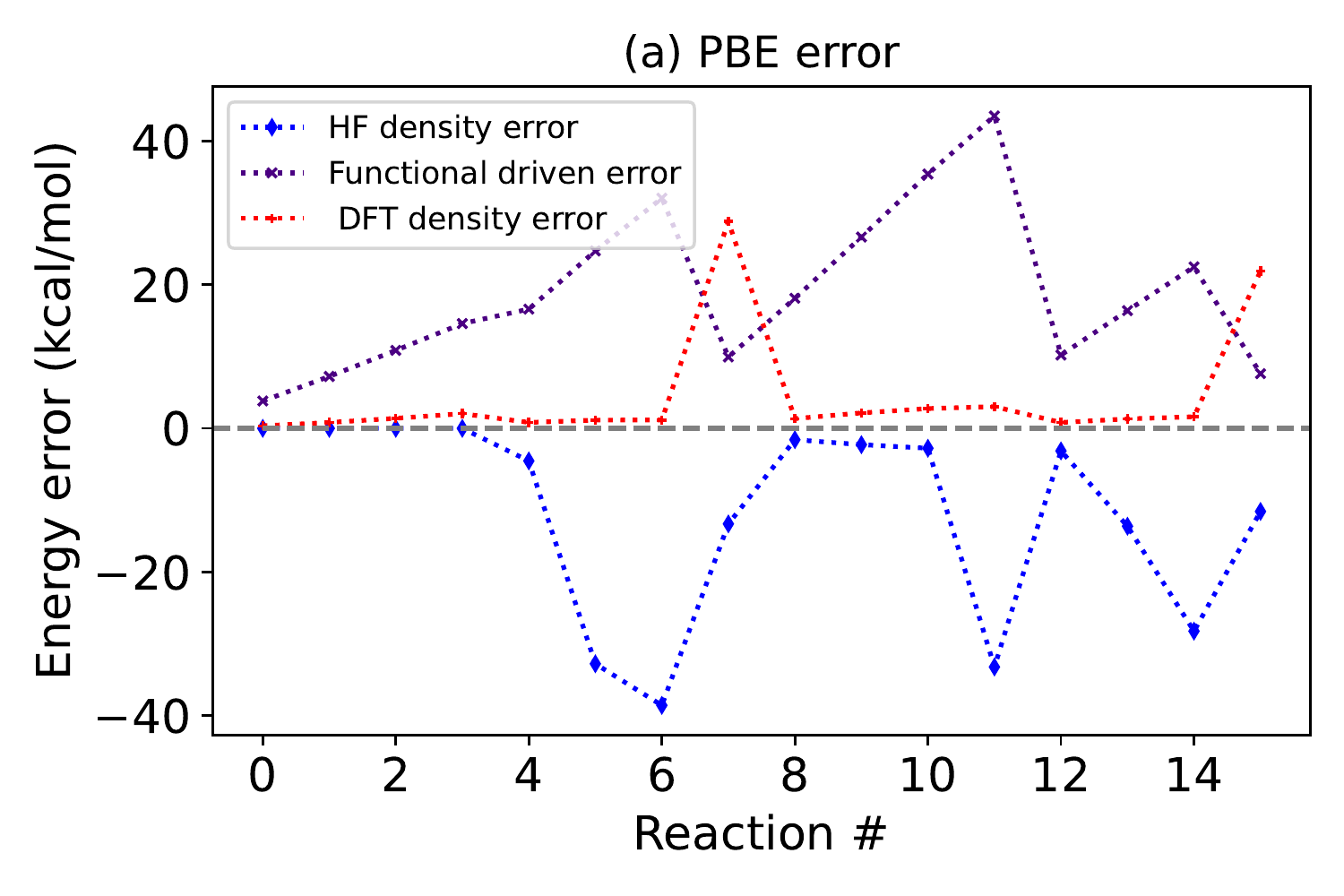}
    \label{fig:SIEtotal_pbe_error}
  \end{subfigure}
  \begin{subfigure}[b]{0.6\linewidth}
    \includegraphics[width=\linewidth]{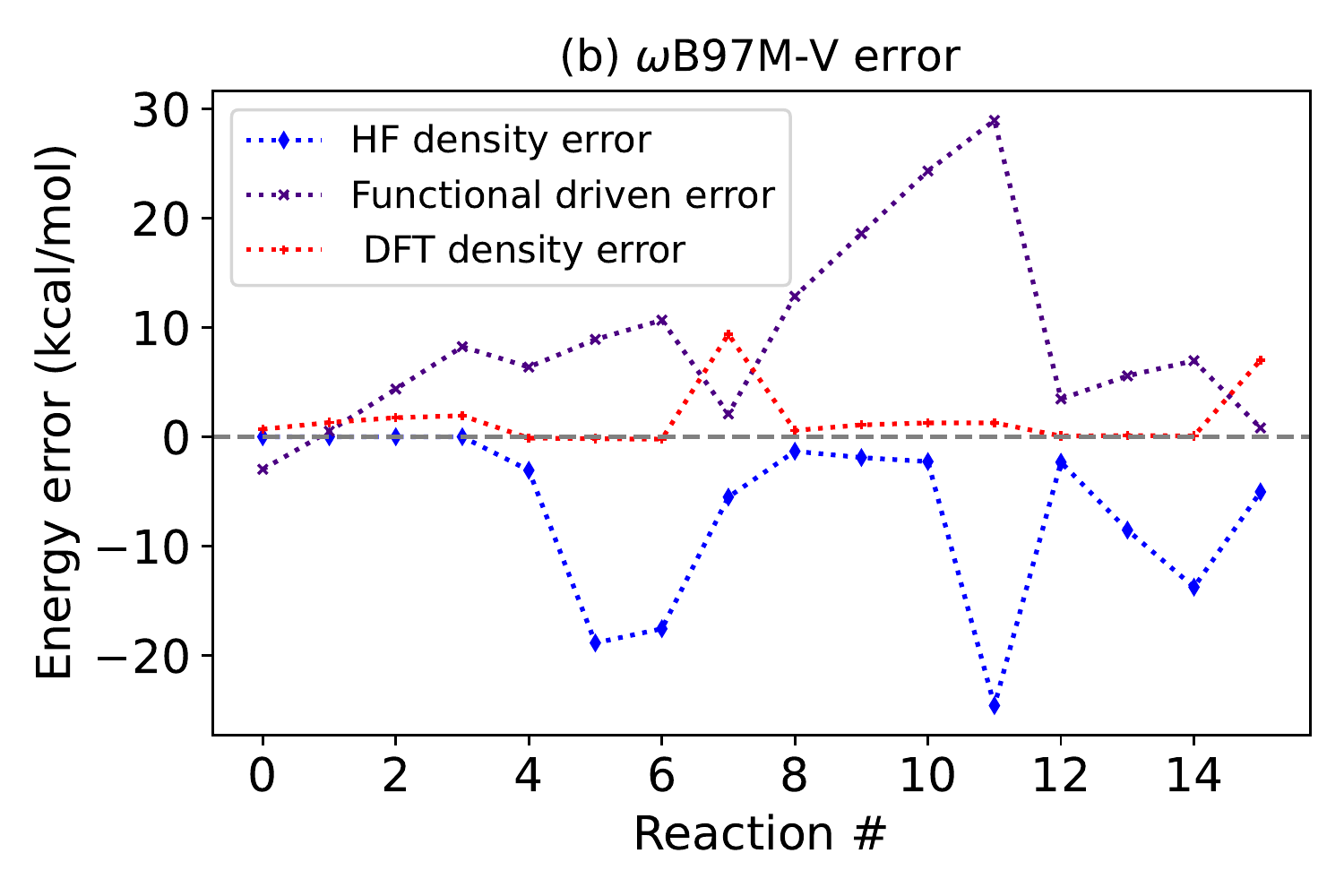}
    \label{fig:SIEdft_densityerror}
  \end{subfigure}
  \\
\caption{SIE4x4 set error decomposition comparing the case of PBE  and $\omega$B97M-V. Similar trends appear between the two, notably large functional errors and comparably small density driven errors  }
\label{fig:SIEerrordecomp}
\end{figure}

Referring to \cref{fig:SIEerrordecomp}, we conclude that functional-driven errors for these difficult SIE-sensitive cases are large, although they get smaller for the higher rung functionals. Use of HF densities introduces additional density-driven error in comparison to use of the self-consistent DFT density for almost all of the 16 data points, which is the cause of the HF-DFT improvements (i.e. partial error cancellation again). However functional-driven errors cannot be addressed by using better-than-HF densities. 

\begin{figure}[htbp]
    \centering
    \includegraphics[width=0.7\linewidth]{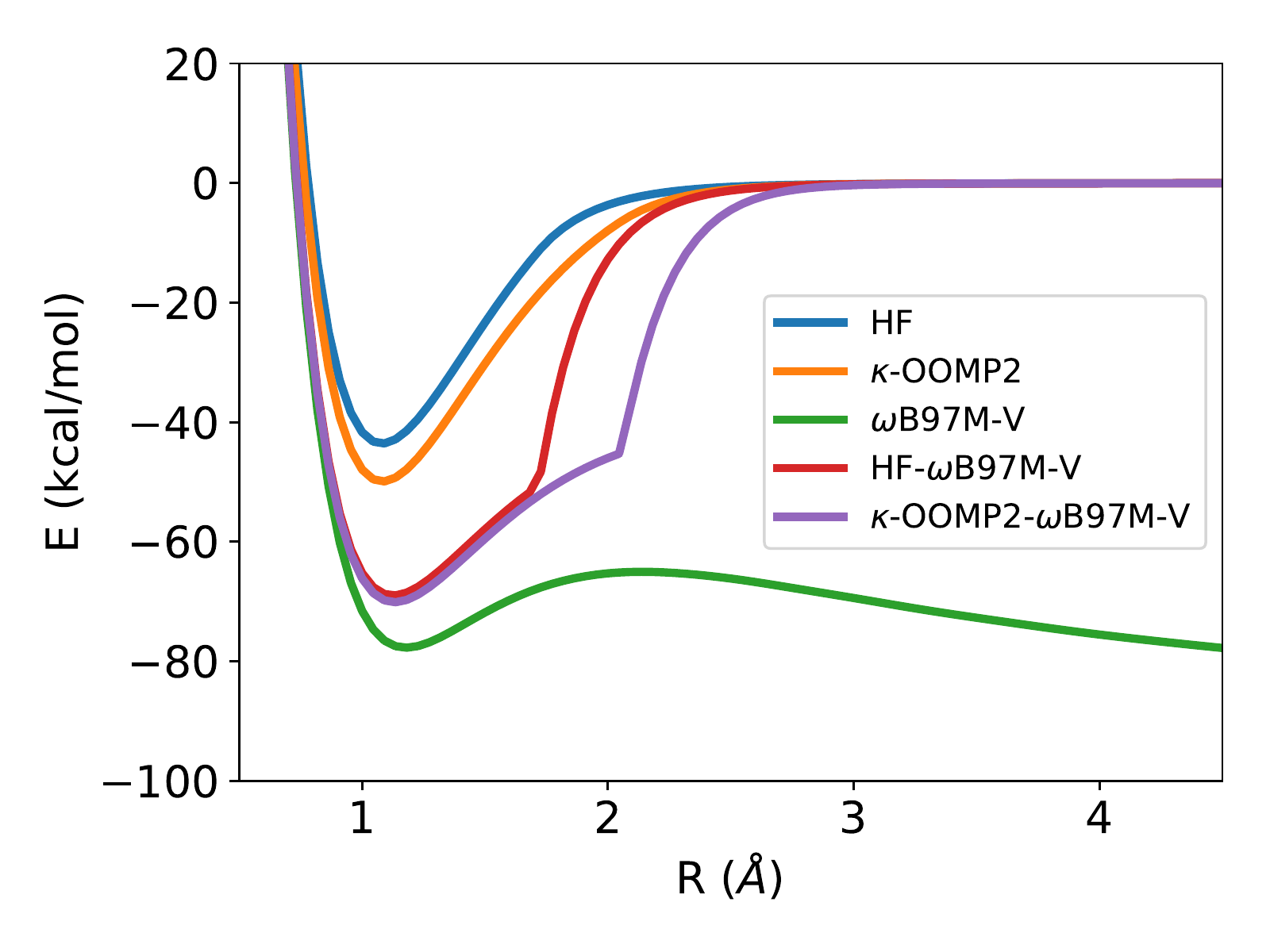}
    \caption{PES in kcal/mol for He$_2^{+}$. Self-consistent use of a good rung 4 functional, $\omega$B97M-V, gives large errors at equilibrium that increase rapidly with $R$, yielding 2 He$^{\frac{1}{2}+}$. Using the same functional with non-selfconsistent densities (HF and $\kappa$-OOMP2) that yield charge-localized products (He$^+$ + He) corrects the error at dissociation but the PESs show a first derivative discontinuity. }
    \label{fig:he2+stretch}
\end{figure}

Finally, we present a more thorough analysis of the He$_2^+$ bond stretch - responsible for 4 datapoints in the SIE4x4 dataset. The potential energy surface (PES) as a function of the bond length computed with the $\omega$B97M-V functional and several density corrected variants is given in \cref{fig:he2+stretch}. $\omega$B97M-V, which includes 100$\%$ exact exchange at long range, fails to bind the molecule due to self interaction error, while HF and $\kappa$-OOMP2 give qualitatively accurate pictures. Near the equilibrium bond length, non-selfconsistent use of the HF and $\kappa$-OOMP2 orbitals with the $\omega$B97M-V gives curves that  largely resemble the $\omega$B97M-V. Density-driven error is not zero, as the $\kappa$-OOMP2 densities evidently removes a small amount of SIE as these curves are shifted up about 10 kcal/mol at the equilibrium bond length.

At the generalized Coulson-Fischer point where spontaneous charge localization begins to occur, the HF and $\kappa$-OOMP2 solutions start to polarize, smoothly localizing the unpaired electron onto a single atom. Note the distinctly different values of $R_\text{C-F}$ for HF and $\kappa$-OOMP2 orbitals: the HF orbitals break symmetry at shorter $R$ due to neglect of electron correlation. This localization removes SIE in the HF-$\omega$B97M-V and $\kappa$-OOMP2-$\omega$B97M-V methods, leading to a rapid increase in energy for $R>R_\text{C-F}$. Evidently, density corrected DFT can fix the asymptotic behavior of bond stretches, but it cannot fix the large errors around the equilibrium bond lengths, which are evidently functional-driven errors.

Moreover, when using $\omega$B97M-V with non-selfconsistent densities, the connection of the short $R$ delocalized limit and the long $R$ localized limit to yield a continuous PES is associated with a pronounced first derivative discontinuity at $R=R_\text{C-F}$. The result is a qualitatively poor representation of the true PES, despite the correct asymptotic limit. The origin of the kink is the fact that in non-selfconsistent DFT, the energy depends on first order changes, $\delta \boldsymbol{\theta}$, in the orbitals, because $\partial E / \partial \boldsymbol{\theta} \ne \mathbf{0}$. At $R=R_\text{C-F}$, the orbitals exhibit a first derivative discontinuity, and therefore so too does the energy derivative $d E / d R$, just like \cref{eq:responsedipole}. For orbitals that are variationally optimized, by contrast, the variational theorem states that the energy depends on $\delta \boldsymbol{\theta}^2$, which accounts for the visually smooth curves despite the first derivative orbital discontinuity.

\section*{Conclusion}
In conclusion, we investigated the use of $\kappa$-OOMP2 densities within the DC-DFT framework rather than the more traditional HF density. We surprisingly found that this higher quality density degraded results for barrier heights, halogen bonding, and bond dissociations. 
This appears to contradict the assertion that density driven errors are the source of bad energetics for systems prone to self-interaction error, motivating an analysis of these errors through the lens of DC-DFT. We found that functional driven errors are the dominant error source, wherein even the exact density can do little to remedy large self-interaction errors within a given density functional approximation.
We attribute the success of HF-DFT type methods to favorable error cancellation between overly delocalized density functional approximations and overly localized Hartree-Fock densities.
HF-DFT and by extension DC(HF)-DFT are vulnerable to cases where the creation of density driven error may overcompensate or where this cancellation does not occur. It is therefore advised to use caution when using DC(HF)-DFT. For a systematic improvement, a proxy for the exact density such as $\kappa$-OOMP2 densities have been demonstrated to provide more accuracy than lower-scaling density functionals while also reducing the total absolute error in the calculation. 

We additionally presented a metric to quantify the degree of error cancellation present in HF-DFT through $E_{\text{TA}}$. We believe this metric will be helpful in distinguishing between ``true" improvement and fortuitous cancellation of error within density corrected DFT methods.

\section*{Acknowledgments}
This work was supported by the Director, Office of Science, Office of Basic Energy Sciences, of the U.S. Department of Energy under Contract No. DE-AC02-05CH11231. A.R. acknowledges funding from the National Science Foundation under award number DGE 1752814. 

\section*{Author Declarations}
Martin Head-Gordon is a part-owner of Q-Chem, which is
the software platform used to perform the calculations described in this work.

\bibliography{references}

\begin{tocentry}
\includegraphics[width=1\linewidth]{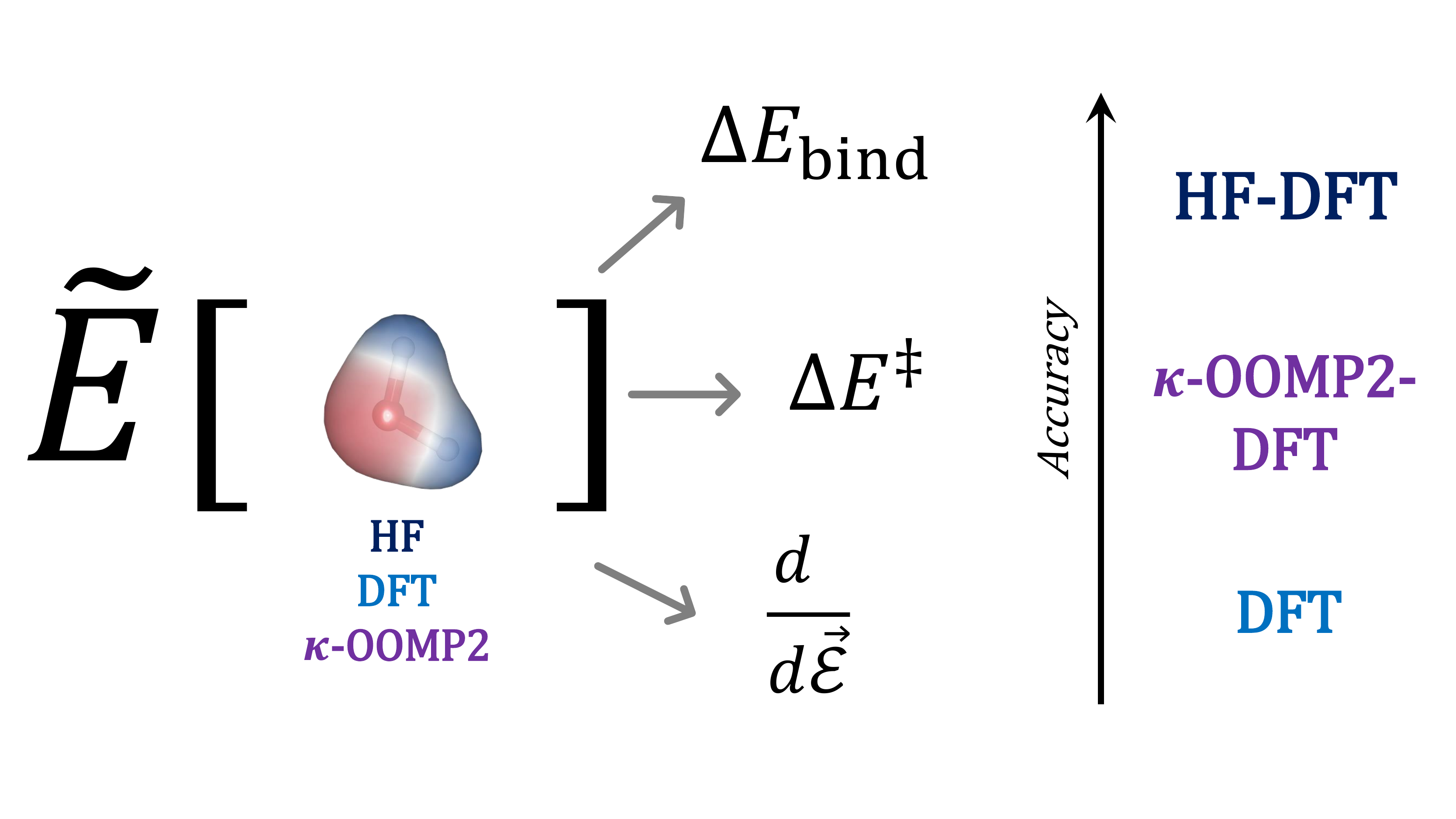}
\end{tocentry}

\end{document}